\documentstyle[aps,prb,epsf,multicol]{revtex}

\newcommand{\bleq}{\ifpreprintsty
                   \else
                   \end{multicols}\vspace*{-3.5ex}{\tiny 
                   \noindent\begin{tabular}[t]{c|}
                   \parbox{0.493\hsize}{~} \\ \hline \end{tabular}}
                   \fi}
\newcommand{\eleq}{\ifpreprintsty
                   \else
                   {\tiny\hspace*{\fill}\begin{tabular}[t]{|c}\hline
                    \parbox{0.49\hsize}{~} \\ 
                    \end{tabular}}\vspace*{-2.5ex}\begin{multicols}{2}
                    \fi}
\newcommand{\bcols}{\ifpreprintsty\else\begin{multicols}{2}\fi}
\newcommand{\ecols}{\ifpreprintsty\else\end{multicols}\fi}

\begin{document}
\draft
\title{Weak localization and conductance fluctuations of a 
chaotic quantum dot with tunable spin-orbit coupling}
\author{P.\ W.\ Brouwer,$^a$ J.\ N.\ H.\ J.\ Cremers,$^b$ 
B.\ I.\ Halperin$^b$}
\address{$^{a}$Laboratory of Atomic and Solid State Physics,
Cornell University, Ithaca, NY 14853-2501\\
$^{b}$Lyman Laboratory of Physics, Harvard University, Cambridge MA 
02138\\
{\rm \today}
\medskip \\ \parbox{14cm}{\rm
In a two-dimensional quantum dot in a GaAs heterostructure, 
the spin-orbit scattering rate is substantially
reduced below the rate in a bulk two-dimensional electron gas 
[B.I.\ Halperin {\em et al}, Phys.\ Rev.\ Lett.\
{\bf  86}, 2106 (2001)].
Such a reduction can be
undone if the spin-orbit coupling parameters acquire a spatial
dependence, which can be achieved, e.g., by a metal gate covering
only a part of the quantum dot. 
We calculate the effect of such spatially
non-uniform spin-orbit scattering
on the weak localization correction and the universal
conductance fluctuations of a chaotic quantum dot coupled to
electron reservoirs by ballistic point contacts, in 
the presence of a magnetic field parallel to the plane of the
quantum dot. 
\medskip\\
PACS numbers: 73.23.-b, 73.20.Fz, 72.10.Bg, 72.20.-i}}
\maketitle

\bcols

In disordered metals, interference of time-reversed
trajectories leads to a small negative correction to the 
conductivity, known as weak 
localization.\cite{mesoreview1,SimonsAltshuler} With
strong spin-orbit scattering, the effect of such interference 
is opposite, causing weak antilocalization, a
positive correction to 
the conductivity.\cite{Hikami,Bergmann} 
Both interference corrections are suppressed when
time-reversal symmetry is broken by a magnetic field.
The same phenomena are observed in a
two-dimensional electron gas, such as is formed in GaAs
heterostructures.
While spin-orbit scattering
in metals is largely due to scattering from the 
metal ions or from impurities, in a GaAs heterostructure, 
spin-orbit effects mainly arise from the asymmetry of the potential
creating the quantum well (Rashba term), as well as from the lack of 
inversion
symmetry which may occur in the crystal structure of the material
forming the heterostructure (Dresselhaus term). 

Recently, it has become possible to study spin-orbit 
scattering in finite size systems, such as
metal grains and semiconductor quantum 
dots.\cite{Ralph,Davidovic,Folk,Hackens} In the universal
regime, where all relevant 
time scales (spin-orbit time $\tau_{\rm so}$, inverse level
spacing/broadening) 
are much larger than the electron transit time $\tau_{\rm erg}$,
such systems can be described using random-matrix
theory.\cite{Beenakker97}
Even though $\tau_{\rm so} \gg \tau_{\rm 
erg}$ in the universal regime, spin-orbit scattering may still have
a significant effect on wavefunctions and
transport properties if $\tau_{\rm so}$
is comparable to the inverse level spacing or level broadening, 
respectively.

For metal grains, the spin-orbit Hamiltonian $H_{\rm so}$
is modeled by a random hermitian matrix with symplectic 
symmetry,\cite{Halperin} the same symmetry as 
in the case of bulk disordered metals.
However, for GaAs quantum dots, 
the situation is more complicated: both the symmetry
of the random matrix representing $H_{\rm so}$ and the 
spin-orbit time are different from
the case of a bulk two-dimensional electron gas.\cite{Stern,AF} 
The complications arise from the special form
of the spin-orbit Hamiltonian $H_{\rm so}$ in GaAs,
\begin{equation}
  H_{\rm so} = {1 \over 2 m} \left( {p_y \sigma_x \over \lambda_y} -
  {p_x \sigma_y \over \lambda_x} \right)
  + {\cal O}(p^3),
  \label{eq:Hso}
\end{equation}
where $\lambda_x$ and $\lambda_y$
are length scales describing the spin-orbit scattering strength
in a GaAs 
heterostructure\cite{RashbaDresselhaus} 
and $\vec \sigma$ are the Pauli matrices. 
The structure of $H_{\rm so}$ is that of a ``non-Abelian vector
potential'', coupled to the electron's spin.\cite{MathurStone}
As this ``vector potential'' has no spatial dependence,
and hence no ``flux'', a suitable
gauge transformation removes the spin-orbit 
scattering term from the Hamiltonian up to corrections of order
$L/\lambda_{x,y}$ which arise due to the 
non-Abelian nature of the ``vector potential'' of Eq.\ (\ref{eq:Hso}).
(Here $L$ is the size of the dot; In the universal regime
one has $L \ll \lambda_{x,y}$.\cite{AF})  
As a result, the spin-orbit scattering 
time is increased by
a large factor\cite{Stern,AF} 
$\sim \lambda_x \lambda_y/L^2 \sim 
\tau_{\rm so}^{\infty}/\tau_{\rm erg}$
over its value 
$\tau_{\rm so}^{\infty} \sim 2\lambda_x \lambda_y/ \ell v_F$
in a bulk two-dimensional electron gas 
with Fermi velocity $v_F$ and mean free path $\ell$
equal to that in the dot, or with $\ell \sim L$ for the case of a ballistic dot. 
Moreover, as was shown in Ref.\ \onlinecite{AF}, 
the symmetry of the transformed spin-orbit scattering term is
not symplectic, but unitary.\cite{unitary}

\begin{figure}
\epsfxsize=0.6\hsize
\hspace{0.2\hsize}
\epsffile{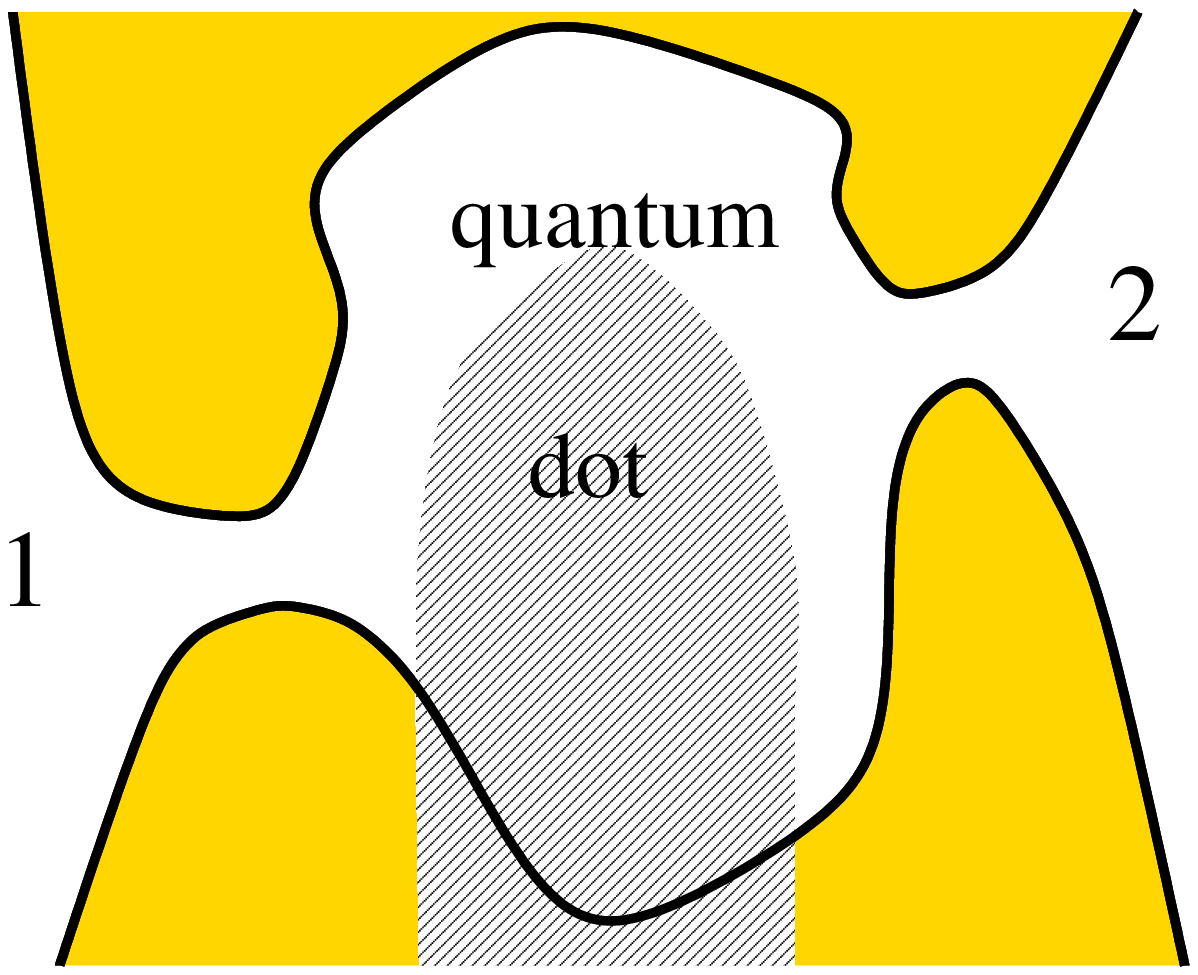}
\refstepcounter{figure}
\label{fig:dot} 
\bigskip

{\small \rm FIG.\ \ref{fig:dot}.
Schematic drawing of a quantum dot
with a metal gate (hatched) over part of the dot. The role of
the gate is to change the asymmetry of the potential of the
quantum well beneath it, and hence the spin-orbit parameters 
$\lambda_x$ and $\lambda_y$.}
\end{figure}

In this paper we investigate the case where the spin-orbit coupling
parameters $\lambda_x$ and $\lambda_y$ are not constant throughout 
the quantum dot. Experimentally, such
a situation could be created with the help of a metal gate parallel 
to the two-dimensional electron gas that changes the asymmetry of the 
quantum well.\cite{Shayegan,Marcus} If the metal gate covers only
a part of the quantum dot, as is shown schematically in Fig.\
\ref{fig:dot}, the translational invariance of $H_{\rm so}$ is lifted.
Hence, the spin-orbit scattering can no longer be gauged away to 
leading order
in $L/\lambda_x$, $L/\lambda_y$. 
In other words, the ``non-Abelian vector potential''
in Eq.\ (\ref{eq:Hso}) now represents a nonzero ``flux''. 
The consequence is a significant increase of the spin-orbit
scattering rate and a restoration of the
symplectic symmetry of $H_{\rm so}$.
Thus, a metal gate that changes the asymmetry of the quantum well in
only part of the quantum dot has a fundamentally different effect than
a metal gate that changes the quantum well potential uniformly
throughout the dot. A gate that covers only part of the dot 
is clearly the more effective tool to tune the spin-orbit scattering 
rate.

As an example, let us consider a quantum dot for which the
spin-orbit
scattering is due to the asymmetry of the potential well 
(Rashba term) only,
so that $|\lambda_x| = |\lambda_y| = 
\lambda$.\cite{RashbaDresselhaus} 
If there is a metal gate over
half of the dot as in Fig.\ \ref{fig:dot}, the spin-orbit 
scattering rate may take two different values $\lambda^{-1} =
\bar \lambda^{-1} \pm \case{1}{2}\lambda_{\rm s}^{-1}$ in the two 
halves of the dot.
As discussed in Refs.\ \onlinecite{Stern,AF}, the spatially uniform 
component of $H_{\rm so}$ leads to a unitary perturbation of
the Hamiltonian, with a characteristic time 
\begin{equation}
\tau_{\rm so}^{\rm u} \sim \tau_{\rm erg}(\bar \lambda/L)^4
  \sim (\tau_{\rm so}^{\infty})^2/\tau_{\rm erg}.
  \label{eq:tau1}
\end{equation} 
The spatially varying component
gives rises to a symplectic perturbation of the Hamiltonian, with a 
characteristic time that can be estimated as the time to
accumulate a ``flux quantum'' from the ``vector potential'' in
Eq.\ (\ref{eq:Hso}),
\begin{equation}
  \tau_{\rm so}^{\rm s} \sim
   \tau_{\rm erg}(\lambda_{\rm s}/L)^2.
  \label{eq:tau2}
\end{equation}
If $\lambda_{\rm s} \sim \lambda$,
$\tau_{\rm so}^{s}$ becomes 
comparable to $\tau_{\rm so}^{\infty}$, the spin-orbit scattering time 
in a bulk two-dimensional electron gas. 

We now present a quantitative calculation of how such
a tunable spin-orbit scattering time affects the quantum 
interference corrections to the conductance:
the weak (anti)localization correction 
$\langle \delta G \rangle$ and the conductance autocorrelation
function $\mbox{cov}[G(\vec B), G(\vec B')]$, where $\vec B$
is a magnetic field. 
Our work extends previous works of Efetov \cite{Efetov} and Frahm
\cite{Frahm} for the magnetic-field dependent quantum interference
corrections without spin-orbit scattering or a parallel magnetic
field. The effect of the spatially uniform component of $H_{\rm so}$
on the weak localization correction was calculated in Ref.\
\onlinecite{AF}.

We consider a quantum dot
coupled to two electron reservoirs (labeled $1$ and $2$),
via ballistic point contacts that have
$N_1$ and $N_2$ channels each. We assume that the electron
motion in the quantum dot is chaotic, so that random matrix
theory can be used to calculate the conductance distribution
in the universal regime $g \mu_B B, \hbar/\tau_{\rm so} \ll 
\hbar/\tau_{\rm erg}$.\cite{Beenakker97} (Here $g$ is the
electron $g$-factor and $\mu_B$ the Bohr magneton.)
The quantum dot is described in terms of its scattering matrix
$S$, which, for particles with spin, is a $N \times N$ unitary 
matrix of quaternions, $N = N_1 + N_2$. 
Quaternions are $2 \times 2$ matrices
with special rules for transposition and complex 
conjugate.\cite{Mehta} 
Starting point of the calculation is the Landauer
formula for the two-terminal conductance $G$ of the
quantum dot at zero temperature,
\begin{equation}
  G = {2 e^2 \over h} {N_1 N_2 \over N} 
  - {e^2 \over h} 
  \mbox{tr}\, S \Lambda S^{\dagger} \Lambda\,
  ,
  \label{eq:Landauer}
\end{equation}
where the diagonal matrix $\Lambda$ has elements
$$
  \Lambda_{jj} = \left\{ \begin{array}{ll}
  N_{2}/N, & j=1,\ldots,N_1, \\
  -N_{1}/N, & j=N_1+1,\ldots,N. \end{array} \right.
$$

In order to find the average and variance of the conductance $G$
it is sufficient to compute the average
$$
  \langle S_{kl;\mu\nu}(\varepsilon,\vec B) 
  S_{k'l';\mu'\nu'}(\varepsilon',\vec B')^* \rangle
$$
in the presence of spin-orbit scattering 
and for arbitrary values of
the magnetic field $\vec B$ and Fermi energy $\varepsilon$.
(Roman indices refer to the propagating channels in the leads, greek
indices refer to spin.)
In a random-matrix approach, the statistical properties of 
the scattering matrix $S$ can either be calculated from a 
hermitian random matrix
that represents the Hamiltonian of the quantum dot, or from
a random unitary matrix.\cite{Beenakker97} Here we use
the latter approach; equivalence of the two approaches, including
the dependence on an external parameter, was shown in 
Ref.\ \onlinecite{WavesRM}. The $N \times N$
matrix $S$ is written as\cite{WavesRM}
\begin{equation}
  S = P U (1 - Q^{\dagger} R Q U)^{-1} P^{\dagger}, \label{eq:SU}
\end{equation}
where $U$ is an $M \times M$ random
unitary symmetric matrix taken from Dyson's
circular orthogonal ensemble\cite{Beenakker97}
and $R$ is a unitary matrix of size $M-N$. The $N \times M$ matrix $P$
and the $(M-N) \times M$ matrix $Q$ are projection matrices with $P_{ij}
= \delta_{i,j}$ and $Q_{ij} = \delta_{i+N,j}$. The quaternion elements
of the matrices $U$, $P$, and $Q$ are all proportional to the 
$2 \times 2$ unit matrix $\openone$. The matrix $R$ is given by
\begin{equation}
  R(\varepsilon,\vec B,\tau_{\rm so}) = 
  \exp\left[{i 2 \pi \over M \Delta} 
  (\varepsilon - H'(\vec B,\tau_{\rm so}))\right],
\end{equation}
where $\Delta$ is the mean level spacing of the dot and
$H'$ is an $(M-N)$ dimensional quaternion matrix generating the
perturbations to the dot Hamiltonian that correspond to the magnetic
field and the spin-orbit scattering,
\begin{eqnarray}
  H' &=&  
       {\mu_B g \over 2}
   \vec B \cdot \vec \sigma + i {x \Delta \over 2 \pi} X \openone 
%  \nonumber \\ && \mbox{}
  + i \sqrt{{ \hbar \Delta \over 2 \pi \tau_{\rm so}}} (A_1 \sigma_x +
  A_2 \sigma_y). \nonumber \\
  \label{eq:dH}
\end{eqnarray}
Here $A_j$ ($j=1,2$) and $X$ are real antisymmetric matrices of 
dimension $M-N$, with $\mbox{tr}\, A_i A_j^{\rm T} = M^2 \delta_{ij}$
and $\mbox{tr}\,
X X^{\rm T} = M^2$.\cite{footso}
The symmetry of the spin-orbit term in Eq.\ (\ref{eq:dH}) is
chosen in accordance with Eq.\ (\ref{eq:Hso}), taking into account
that the spin-orbit Hamiltonian has symplectic symmetry once the
coupling parameters $\lambda_x$ and $\lambda_y$ depend on position.
In Eq.\ (\ref{eq:dH}) the orbital and Zeeman effects of the magnetic
field have been separated. The first term
describes the Zeeman coupling
to the spin of the electrons. The second term models the
orbital effect, where $x$ is related to the perpendicular
component of the magnetic field,
$$
  x^2 = {c e^2 L^4 B_{\perp}^2 / (\hbar 
 \tau_{\rm erg} \Delta) }.
$$
$c$ being a numerical coefficient of
order unity.\cite{c} 
At the end of the calculation, the limit $M \to \infty$ should be
taken. 

We now describe our calculation, which was done to leading order in
$1/N$. Corrections for finite $N$ are discussed at the end of this
paper.
To leading order in $1/M$ and $1/N$,
it is sufficient to consider
the elements of $U$ as random Gaussian variables with zero mean
and with variance $\langle U_{ij} U_{kl}^* \rangle = M^{-1}
(\delta_{ik} \delta_{jl} + \delta_{il} \delta_{jk})$.\cite{BB-jmp}
We then expand Eq.\ (\ref{eq:SU}) in powers of $U$ and perform the
Gaussian averages to leading order in $1/M$. We thus find
\begin{eqnarray}
  && \langle S_{kl;\mu\nu}(\varepsilon,\vec B) 
  S_{k'l';\mu'\nu'}(\varepsilon',\vec B')^* \rangle
  \nonumber \\ && \mbox{} =
%  &=&
  \delta_{kk'} \delta_{ll'}
  D_{\mu\nu;\nu'\mu'}
%   \nonumber \\ && \mbox{}
  +
  \delta_{kl'} \delta_{lk'}
  ({\cal T} C {\cal T})_{\mu\nu;\mu'\nu'},
  \label{eq:result}
\end{eqnarray}
where, in tensor notation,
\begin{eqnarray*}
  D &=&
  \left( M \openone \otimes \openone - 
  \mbox{tr}\, R \otimes R'^{\dagger} \right)^{-1},
  \\
  C &=&
  \left( M \openone \otimes \openone - 
  \mbox{tr}\, R \otimes R'^{*} \right)^{-1}.
\end{eqnarray*}
Here $R'^*$ is the quaternion complex conjugate of $R'$, ${\cal T} = 
\openone \otimes \sigma_2$, and the tensor multiplication should be 
understood as ``backwards
multiplication'' for the second matrix, i.e., with the 
multiplication rules
$$
  (\sigma_i \otimes \sigma_j) (\sigma_{i'} \otimes \sigma_{j'})
  = (\sigma_i \sigma_{i'}) \otimes (\sigma_{j'} \sigma_{j}).
$$
The two contributions $C$ and $D$ are the equivalents of cooperon and
diffuson in the conventional diagrammatic perturbation 
theory.\cite{SimonsAltshuler}

Taking the limit $M \to \infty$ and defining a dimensionless
magnetic field $b = \pi g \mu_B B/\Delta$ and spin-orbit scattering
rate $a^2 = 2 \pi \hbar/\tau_{\rm so} \Delta$, we find
\begin{eqnarray}
  D^{-1} 
  &=& N_D (\openone \otimes \openone)
%  \nonumber \\ && \mbox{}
      + i \vec b \cdot (\vec \sigma \otimes \openone)
      - i \vec b' \cdot (\openone \otimes \vec \sigma)
  \nonumber \\ && \mbox{}
      + 2 a^2 (\openone \otimes \openone) 
      - a^2 (\sigma_x \otimes \sigma_x +
             \sigma_y \otimes \sigma_y),
  \label{eq:D}
  \\
  C^{-1} &=& N_C
      (\openone \otimes \openone)
%  \nonumber \\ && \mbox{}
      + i \vec b \cdot (\vec \sigma \otimes \openone)
      + i \vec b' \cdot (\openone \otimes \vec \sigma)
  \nonumber \\ && \mbox{}
      + 2 a^2 (\openone \otimes \openone) 
      - a^2 (\sigma_x \otimes \sigma_x +
             \sigma_y \otimes \sigma_y),  \label{eq:C}
\end{eqnarray}
where $N_D$ and $N_C$ are given by
\begin{eqnarray}
  N_D &=& N - 2 \pi i (\varepsilon - \varepsilon')/\Delta + (1/2) (x-x')^2, 
  \label{eq:ND} \\
  N_C &=& N - 2 \pi i (\varepsilon - \varepsilon')/\Delta + (1/2) (x+x')^2.
  \label{eq:NC}
\end{eqnarray}

\begin{figure}[h]
\epsfxsize=0.99\hsize
\epsffile{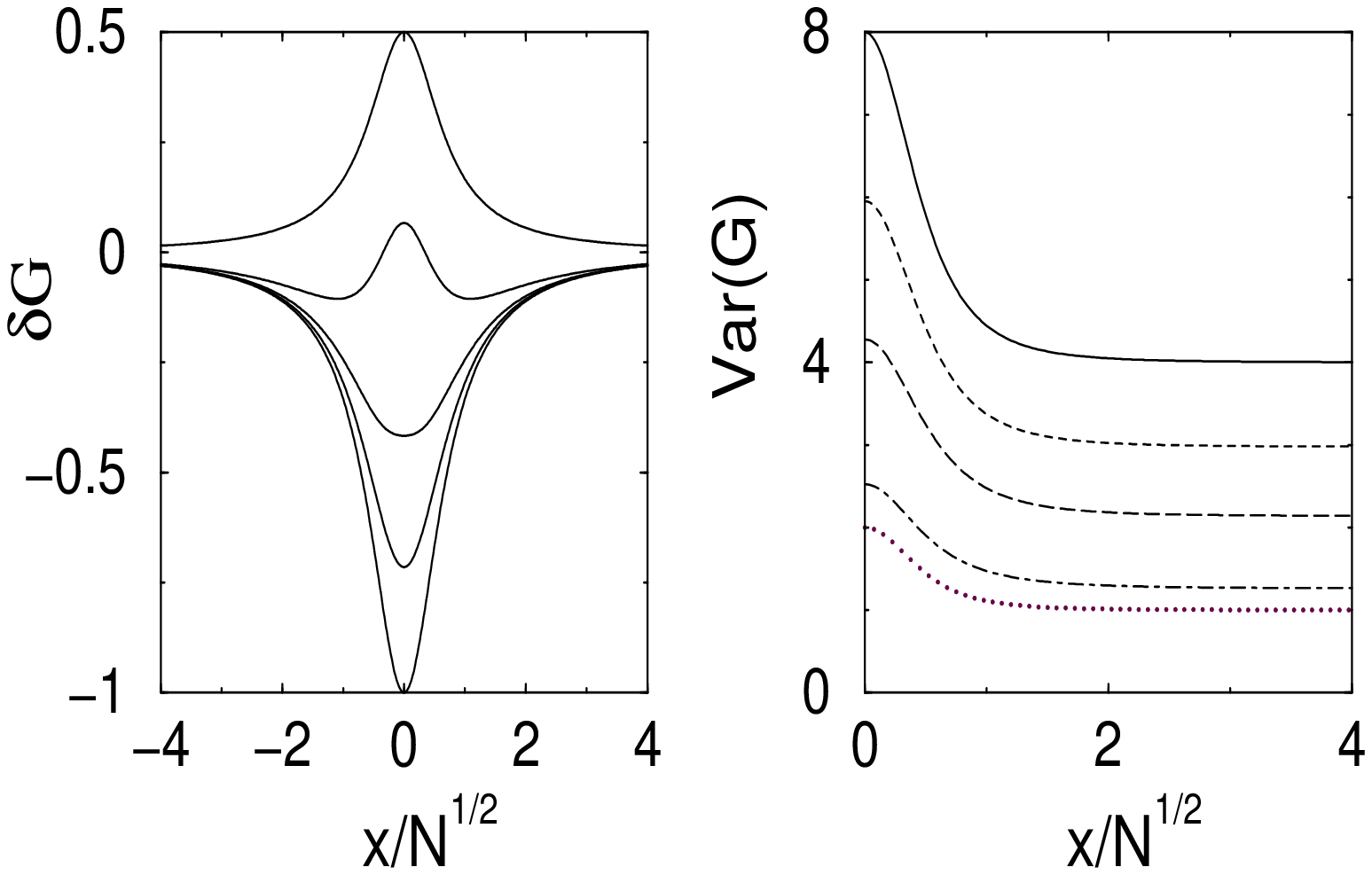}
\refstepcounter{figure}
\label{fig:2}
{\small FIG.\ \ref{fig:2}. Left panel: Weak localization correction
$\langle \delta G \rangle$ as a function of the dimensionless
perpendicular magnetic field $x$, for $a=0$ (bottom curve), $a=0.3$,
$a=0.5$, $a=1.0$, and $a \to \infty$ (top curve). Right panel: $\mbox{var}\,
G$ versus $x$ for the same values of $a$ ($a=0$ is top curve, $a \to
\infty$ is bottom curve). In both cases, there
is no parallel component of the magnetic field and $G$ is measured
in units of $(e^2/h)N_1 N_2/N^2$.}
\end{figure}

We now set $\vec b = b \hat x$, $\vec b' = b' \hat x$, take
the inverses in Eqs.\ (\ref{eq:D}) and (\ref{eq:C}), 
and calculate the average and covariance of the conductance $G$
from Eq.\ (\ref{eq:Landauer}). For $\langle G \rangle$, we find
\bleq
\begin{eqnarray}
  \langle G(\varepsilon,x,b) \rangle &=& 
  {2 e^2 \over h} {N_1 N_2 \over N_1 + N_2}
  \left[1 - {1 \over 2} \left( {1 \over N_C + 2 a^2} + {1 \over N_C + 4 a^2}
%  \right. \nonumber \\ && \left. \mbox{}
  - {2 a^2 \over N_C (N_C+2 a^2) + 4 b^2} \right) \right],
  \label{eq:deltaG1}
\end{eqnarray}
where $N_C = N + 2 x^2$, as follows from Eq.\ (\ref{eq:NC}) with 
$\varepsilon = \varepsilon'$, $x = x'$.
To calculate the zero temperature
conductance fluctuations, it is sufficient to know
the two-point correlator (\ref{eq:result}) to leading order in $1/N$. 
(Contributions from higher-order correlators vanish since
they contain a factor $\mbox{tr}\, \Lambda = 0$.\cite{Argaman}) 
We then find
\begin{eqnarray}
  \mbox{cov}\,[G(\varepsilon,x,b),G(\varepsilon',x',b')] &=&
  \left( {e^2 \over h} \right)^2 \left( {N_1 N_2 \over N_1 + N_2} \right)^2
  (F_D + F_C), \label{eq:varG2}
\end{eqnarray}
where
\begin{eqnarray*}
  F_D &=& {2 (b - b')^2 + 2 |N_D +a^2|^2 + 2 a^4
  \over |(b - b')^2 + N_D (2a^2 + N_D)|^2} +
  {2 a^4 + 2 (b+b')^2 + 2 |N_D + 3 a^2|^2 \over
  |(b + b')^2 +  (N_D + 4 a^2)(N_D + 2 a^2)|^2},
\end{eqnarray*}
\eleq
\noindent
$F_C$ is obtained from $F_D$ by the substitution $x' \to -x'$,
$b' \to -b'$, and $N_D \to N_C$, and $N_D$ and $N_C$ are given in
Eqs.\ (\ref{eq:ND}) and (\ref{eq:NC}) above.
The conductance fluctuations at finite temperature are obtained
from Eq.\ (\ref{eq:varG2}) by multiplication with the derivatives
of the Fermi function at energies $\varepsilon$ and $\varepsilon'$
and subsequent integration over $\varepsilon$ and $\varepsilon'$.
Equations (\ref{eq:deltaG1}) and (\ref{eq:varG2})
recover the result of Refs.\
\onlinecite{Efetov,Frahm} for the spinless case $a = b = 0$.
Further, one has $\langle \delta G \rangle = 
(e^2/h)(N_1 N_2/N^2)$, $\mbox{var}\, G =
  2(e^2/h)^2 (N_1 N_2/N^2)^2$ if there is no parallel magnetic field,
while spin orbit scattering is strong 
($a^2 \gg N$), in agreement with known results for the
circular symplectic ensemble.\cite{Beenakker97}
(For comparison, 
no positive $\langle \delta G \rangle$ is observed for the
case of a spatially uniform spin-orbit coupling, see Ref.\
\onlinecite{AF}.) Equations (\ref{eq:deltaG1}) and (\ref{eq:varG2})
are illustrated in Fig.\ \ref{fig:2} where we show $\langle \delta
G \rangle$ and $\mbox{var}\, G$ as a function of the perpendicular
magnetic field $x$ for various values of the dimensionless spin-orbit
scattering rate $a$.

In the presence of both strong
spin-orbit scattering ($a^2 \gg N$) and a large parallel field
($b^2 \gg N^2$, $N a^2$), all terms contributing to the cooperon $C$
are suppressed. As a result, there is no weak localization
correction to the conductance,\cite{Maekawa,Edelstein}
$\langle \delta G \rangle = 0$,
and the conductance fluctuations are reduced by an additional factor 
two, $\mbox{var}\, G = (e^2/h)^2 (N_1 N_2/N^2)^2$, see also 
Ref.\ \onlinecite{Stern}. Furthermore, the 
conductance is no longer symmetric under reversal of
the perpendicular magnetic field. This observation is best 
illustrated by the correlator $\langle (G(x) - G(-x))^2 \rangle$, 
which goes to $2\, \mbox{var}\, G$ for perpendicular magnetic field 
strengths $x^2 \gg N$, see Eq.\ (\ref{eq:varG2}).
For comparison, in the presence 
of either spin-orbit scattering {\em or} a parallel field (but not
both), one has $G(x) = G(-x)$ for all $x$.

On a phenomenological level, 
dephasing can be added to the current description
via the voltage probe model of 
B\"uttiker.\cite{Buttiker,BB}
In this model, a fictitious voltage probe 
is attached to the quantum dot.
Electrons escape from the dot into the voltage probe at a
rate $1/\tau_{\phi}$, where $\tau_{\phi}$ is the dephasing time,
and are then reinjected from the voltage probe
without phase memory. The escape into the voltage probe is
described by an imaginary term $i \hbar/2\tau_{\phi}$ in
the Hamiltonian. With the form (\ref{eq:Landauer}) of the Landauer 
formula, the reinjection of 
particles from the voltage probe has no effect on the conductance
to order $N^0$.
Hence, to leading order in $1/N$, inclusion of dephasing
amounts to the replacement
$$
  N_{D,C} \to N_{D,C} + {2 \pi \hbar \over \tau_{\phi} \Delta}
$$
in Eqs.\ (\ref{eq:deltaG1}) and (\ref{eq:varG2}) above.

We would like to thank I.\ L.\ Aleiner, V.\ Falko, Yu.\ B.\
Lyanda-Geller,
and C.\ M.\ Marcus for important discussions. 
This work was supported by the
NSF under grant nos.\ DMR 0086509 and DMR 9981283
and by the Sloan foundation.

\ecols

\begin{references}
\bibitem{RashbaDresselhaus} 
The Hamiltonian (\ref{eq:Hso}) is for a two-dimensional electron
gas formed parallel to the (001) plane of GaAs.
The $x$ and $y$ coordinates are defined along axes with 
crystallographic directions $[110]$ and $[1\bar10]$.
In terms of the rates $\gamma$ and $\eta$
for the Rashba and Dresselhaus terms in the spin-orbit
Hamiltonian (notation of Ref.\ \onlinecite{Stern}), 
one has $\hbar/2\lambda_x = -\eta + \gamma$, 
$\hbar/2\lambda_y = -\eta - \gamma$.
\bibitem{unitary}
A small symplectic perturbation of the Hamiltonian remains, with 
a characteristic time that is a factor $[(L/\lambda_x)^2 +
(L/\lambda_y)^2]^{-1}$ larger than the time scale $\tau_{\rm so}^{\rm u}$
characteristic of the perturbation with unitary symmetry, see 
Ref.\ \onlinecite{AF}.
\bibitem{footso}
We assume that the momentum matrix elements multiplying $\sigma_x$
and $\sigma_y$ in the spin-orbit Hamiltonian (\ref{eq:Hso}) are
equal and independently distributed. This is correct when
$\mbox{var}\, \lambda_x^{-1} = \mbox{var}\, \lambda_y^{-1}$ in 
Eq.\ (\ref{eq:Hso}) and the quantum
dot is roughly circular.
\end{references}
\end{document}